\documentclass[journal]{IEEEtran}

\ifCLASSINFOpdf
\else
   \usepackage[dvips]{graphicx}
\fi
\usepackage{url}

\usepackage{amsmath,graphicx}
\usepackage{epstopdf}
\usepackage{setspace}
\usepackage{amsmath}
\usepackage{amssymb}
\usepackage{stfloats}
\usepackage{amsthm}
\usepackage{subcaption}
\usepackage{amsfonts} 
\usepackage{color}
\usepackage{xcolor}

\usepackage{graphicx}

\usepackage[flushleft]{threeparttable} 
\usepackage{booktabs}
\usepackage{multirow}
\usepackage{upgreek}
\usepackage{bm}
\usepackage{nicefrac}
\usepackage{multirow}
\usepackage{cite}
\usepackage{mathtools}
\usepackage{stmaryrd}
\usepackage[mathscr]{euscript}
\usepackage{array}
\usepackage{mathrsfs}

\usepackage{soul}

\newtheorem{theorem}{Theorem}

\newcommand{\nyq}{\text{Nyq}}

\begin{document}

\title{Linear-Bias Time Encoding for Low-Rate Quantized Representation of Bandlimited Signals}

\author{Anshu Arora*, \IEEEmembership{Student Member, IEEE}, Kaluguri Yashaswini*, and Satish Mulleti \IEEEmembership{Member, IEEE}
\thanks{* Equal contribution}
\thanks{The authors are with the Department of Electrical engineering, Indian Institute of Technology Bombay, India (e-mail: anshuarora2604@gmail.com, yashkaluguri2005@gmail.com, mulleti.satish@gmail.com).}}

\markboth{IEEE Signal Processing Letters, VOL. XXX, 2025}
{Shell \MakeLowercase{\textit{et al.}}: Bare Demo of IEEEtran.cls for IEEE Journals}
\maketitle
\begin{abstract}
Integrate-and-fire time encoding machines (IF-TEMs) provide an efficient framework for asynchronous sampling of bandlimited signals through discrete firing times. However, conventional IF-TEMs often exhibit excessive oversampling, leading to inefficient encoding for signals with smoothly distributed information. This letter introduces a linear-bias IF-TEM (LB-IF-TEM), where the bias dynamically tracks the input signal to maintain a nearly constant integrator input, thereby localizing the firing intervals. The resulting concentrated distribution enables effective non-uniform quantization with reduced distortion. Theoretical analysis establishes explicit bounds on the achievable oversampling range, while experimental results demonstrate that the proposed method attains comparable reconstruction accuracy at significantly lower bitrate than existing IF-TEM variants. The LB-IF-TEM thus provides a low-power, communication-efficient, and analytically tractable framework for time-based signal encoding and reconstruction.
\end{abstract}

\begin{IEEEkeywords}
Integrate-and-fire time encoding machines, asynchronous sampling, low-rate sampling, non-uniform quantization, oversampling control, event-driven sampling.
\end{IEEEkeywords}

\section{Introduction}

Uniform sampling measures signal amplitudes at uniform time intervals, followed by amplitude quantization~\cite{NyquistOriginal,50yearsafternyquist}. Although widely used, this approach requires a global clock, making it power-hungry, hardware-intensive, and prone to electromagnetic interference~\cite{low_power_nonuniform1,low_power_nonuniform2}. Non-uniform sampling methods—such as random sampling~\cite{random_sampling}, level-crossing schemes~\cite{levelcrossingsampling,derivative_level_crossing,sampling_zero_crossing}, and time-encoding machines (TEMs)~\cite{differentTEMS1, differentTEMS2}, mitigate these limitations by encoding timing information instead of amplitudes.

Among these, TEMs are particularly appealing as they encode signals through the timing of events (firings), such as threshold crossings, rather than through sampled amplitudes. Moreover, TEMs allow explicit control over minimum and maximum firing intervals, enabling iterative reconstruction algorithms~\cite{lazar2003time,lazar2004time}. Such timing control is generally unavailable in classical level-crossing schemes.

In the conventional integrate-and-fire TEM (IF-TEM), the input signal is biased to remain positive, integrated, and a firing is generated whenever the integral reaches a fixed threshold. The firing density increases with signal amplitude, ensuring reconstruction when all firing intervals are below the Nyquist interval~\cite{lazar2003time,lazar2004time}. However, this condition leads to oversampling in regions with large amplitudes, increasing both the range of firing intervals and the overall data rate. Since firing intervals are quantized for digital storage or transmission, a wider interval range results in higher quantization error for a given bit budget.

To address this, variable-bias IF-TEMs (VB-IF-TEMs) have been proposed~\cite{AdaptiveBiasIFTEM,VB-IFTEM}, reducing oversampling and the number of firings. For instance, the VB-IF-TEM~\cite{VB-IFTEM} limits the oversampling factor (ratio between the Nyquist interval and minimum firing interval) to about 4.23, significantly lower than that of the conventional IF-TEM. However, this design restricts flexibility, as the firing-rate range cannot be tuned beyond a certain point—limiting applications where dense sampling is desirable.

In this work, we propose the \emph{Linear-Bias Integrate-and-Fire Time Encoding Machine (LB-IF-TEM)}, a modified IF-TEM framework that employs a piecewise linear bias designed to closely track the input signal. The bias dynamically adjusts the integrator input, reducing variance in firing intervals while maintaining perfect reconstruction guarantees. This design offers two key advantages:  
(i)~the concentrated distribution of firing intervals enables efficient low-bit quantization via non-uniform quantization (NUQ), yielding improved rate–distortion trade-offs; and  
(ii)~unlike the VB-IF-TEM, the proposed method allows the oversampling factor to be chosen freely based on system requirements.  

Overall, the LB-IF-TEM combines low-power operation with reduced communication overhead, offering a balanced and efficient framework for time-based signal encoding. While linear biasing has been explored previously~\cite{diff_and_fire}, earlier approaches relied on differentiating the signal and were restricted to finite rate-of-innovation models, unlike the general bandlimited setting considered here.

The rest of the paper is organized as follows. Section~\ref{sec:PF} formulates the problem. Section~\ref{sec:PM} presents the proposed LB-IF-TEM and theoretical analysis. Experimental results are provided in Section~\ref{sec:ER}, and conclusions in Section~\ref{sec:C}.

\section{Problem Formulation}
\label{sec:PF}
We consider the class of bounded, bandlimited signals $\mathcal{B}_{\Omega_0, c}$ defined as
\[
\mathcal{B}_{\Omega_0, c} = \{ f(t) : |f(t)| \leq c,~ F(\omega)=0~\text{for}~|\omega|>\Omega_0 \},
\]
where $F(\omega)$ denotes the Fourier transform of $f(t)$. The corresponding Nyquist sampling rate is $\Omega_0/\pi = 1/T_{\text{Nyq}}$. Furthermore, signals in this class satisfy
\[
|f'(t)| \leq c\Omega_0 = \epsilon,
\]
which bounds their slope by $\epsilon$ \cite{PapoulisSignalAnalysis}.

In a conventional IF-TEM, the signal $f(t)$ is encoded into a sequence of firing times $\{t_n\}_{n\in\mathbb{Z}}$ defined by
\begin{equation}
    \int_{t_{n-1}}^{t_n} \frac{f(t)+b(t)}{\kappa}\, \mathrm{d}t = \Delta, 
    \label{eq:IF_TEM_eq}
\end{equation}
where $\Delta>0$ is the firing threshold, $\kappa>0$ is a scaling constant, and $b(t)$ is a constant positive bias ensuring $f(t)+b(t)>0$. Without loss of generality, $\kappa$ can be absorbed into $\Delta$, and hence we set $\kappa=1$ throughout.

The firing intervals $T_n = t_{n+1} - t_{n}$ are bounded such that
\begin{align}
    T_{\min} \leq T_n \leq T_{\max}, \notag
\end{align}
where the bounds depend on $\Delta$, $b(t)$, and the signal amplitude $c$. The sequence $\{T_n\}$ uniquely determines $f(t)$ provided that $T_{\max} \leq T_{\text{Nyq}}$ \cite{lazar2003time, lazar2004perfect}.

In practice, the firing intervals are quantized for storage or digital processing using either uniform (UQ) or non-uniform quantization (NUQ) \cite{gersho2012vector, hila2022time}. As shown in our previous work \cite{yashaswini2025non}, NUQ achieves lower distortion than UQ for the same bit budget within the IF-TEM framework. Since the quantization error depends on the interval range $T_{\max}-T_{\min}$, in the case of UQ, reducing this range directly improves encoding precision. However, it is more advantageous to have an effective NUQ, hence we would like to have low variance in the firing intervals \cite{UQvsNUQ2}.

In this work, we focus on \emph{range concentration} as the primary design goal and propose a linear-bias strategy that adaptively controls $b(t)$ to localize the variance of firing intervals within a narrower range, thereby improving the quantization efficiency of the IF-TEM.

\section{Design and Analysis of the LB-IF-TEM}
\label{sec:PM}
In this section, we introduce the proposed LB-IF-TEM and show that by enabling bias to closely follow the signal, the interval's range could be minimized. First, we discuss the biased selection and then derive the bounds.

\subsection{Linear bias proposition}
In this work, the bias is a piecewise linear function that keeps the same slope, but gets shifted after every firing. The bias for $t \in [t_n, t_{n+1}]$ is denoted as $b_n(t)$. Bias is called feasible if $f(t)+b_n(t)>0$ for $t \in [t_n, t_{n+1}]$. To derive the linear bias, we consider the bias calculation after firing $t_n$. We assume that within the previous firing interval, $t \in [t_{n-1}, t_{n}]$, the bias was feasible and linear. Next, we discuss the main principle of bias calculation.

Our approach to bias calculation is based on the following characteristics of bandlimited signals. If $f(t_n)$ is known, then by using the fact that $|f^{\prime}(t)|\leq \epsilon$ we have $-\epsilon(t-t_n)+f(t_n)\leq f(t) \leq \epsilon(t-t_n)+f(t_n)$, for $t\geq t_n$. This implies that the signal lies between two straight lines with slopes $\pm \epsilon$. Importantly, the bias will decrease along the straight line with slope $-\epsilon$. Since the bias should ensure that the biased signal is positive, if we choose the bias as $b_n(t) = \epsilon(t-t_n)-f(t_n) + \mu$, where $\mu>0$, then we have that $f(t)+b_n(t)>0$ for $t\geq t_n$.

The approach mentioned previously requires the knowledge of $f(t_n)$. However, an IF-TEM approach measures the signal's averages rather than the instantaneous values. To address the issue, we first demonstrate that the minimum and maximum values of $f(t_n)$ can be determined from the previous TEM measurement. Then, by using the minimum value of $f(t_n)$, we update the bias rule as discussed next.

From the previous time encodings $\{t_n, \, t_{n-1}\}$ and bias $b_{n-1}(t)$, the average value of the signal $\hat{f}_{n-1}$ is determined using \eqref{eq:IF_TEM_eq} as follows:
\begin{align}
    \hat{f}_{n-1} = \frac{1}{T_{n-1}} \int_{t_{n-1}}^{t_n} f(t)\, dt= \frac{\Delta}{T_{n-1}} - \frac{\int_{t_{n-1}}^{t_n} b_{n-1}(t)\, dt}{T_{n-1}}.
    \label{eq:running_fhat}
\end{align}
Since \( f(t) \) is continuous (being band-limited with finite energy), by the Mean Value Theorem, there exists a point \( \xi_n \in (t_{n-1}, t_n) \) such that \( f(\xi_n) = \hat{f}_{n-1} \).
From point \( \xi_n \), the signal can increase or decrease with a maximum slope of \( \pm\epsilon \). By drawing lines with these slopes starting at \( \hat{f}_{n-1} \) at \( \xi_n \), and intersecting them with the vertical at \( t_n \), we determine a range for \( f(t_n) \) (cf. Fig.~\ref{fig:time_analysis}). Specifically, we obtain the bound:
\begin{align}
    f(\xi_n) - (t_n - \xi_n)\epsilon \leq f(t_n) \leq f(\xi_n) + (t_n - \xi_n)\epsilon
    \label{eq:bound_f(tn)}
\end{align}

Now, instead of using the exact value of $f(t_n)$ for bias calculation, if we consider its minimum value, as derived in our previous work \cite{VB-IFTEM}, the bias would still be feasible. Hence, the proposed bias after the firing at $t = t_n$ is 
\begin{align}
b_n(t) =  \epsilon  (t - t_n) -\hat{f}_{n-1} + \epsilon \frac{(t_n - t_{n-1})}{2} + \mu.
\label{eq:linear_bias}
\end{align}
The proposed linear bias is parametrized by the threshold $\Delta$ and the bias shift factor $\mu$, as seen from equations \eqref{eq:running_fhat} and \eqref{eq:linear_bias}. In the following, we discuss how the change in the parameters affects the firing rates and compare them with existing results.

\subsection{Interval Bounds and Parameter Selection}
In this subsection, we characterize the effect of the linear bias on the firing intervals and provide bounds on the oversampling factor. Specifically, we demonstrate that the upper and lower firing bounds can be user-defined, allowing one to choose the parameters accordingly. Our main results are summarized in the following theorem.

\begin{theorem}[Bounds on Firing Intervals Under Linear Bias]
\label{thm:linear-bias-intervals}
Consider signals in the class $\mathcal{B}_{\Omega_0,c}$ sampled using the LB-IF-TEM with a linear time-varying bias of the form in \eqref{eq:linear_bias}. Suppose a user prescribes bounds $0 < \alpha < \beta \leq 1$ on the firing intervals such that
\begin{align}
\alpha\,T_{\nyq} \leq T_n \leq \beta\,T_{\nyq}
\qquad \forall n.
\label{eq:desired_bounds}
\end{align}
Then, the LB-IF-TEM parameters $\Delta$ and $\mu$ must be chosen as
\begin{align}
\Delta &=
\frac{\epsilon\,\alpha\,\beta\,T_{\nyq}^2(\beta+\alpha)}{\beta-\alpha},
\label{eq:lbiftem_Delta_new}\\[4pt]
\mu &=
\frac{\epsilon\,\alpha\,T_{\nyq}(\beta+\alpha)}{\beta-\alpha},
\label{eq:lbiftem_mu_new}
\end{align}
in order to guarantee the bounds in \eqref{eq:desired_bounds}.
\end{theorem}

\begin{figure}[!t]
        \centering
        \includegraphics[width= 2.8 in]{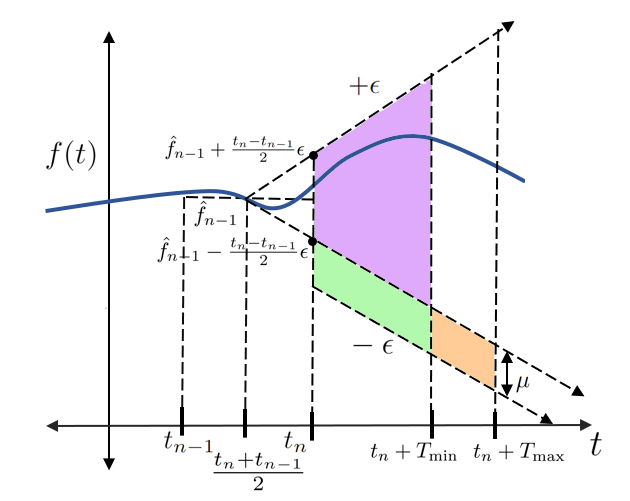}
    \caption{Comparison of areas under the linear bias and the average-time analysis}
    \label{fig:time_analysis}
\end{figure}

\begin{proof}
We prove the interval bounds in two parts by finding $T_{\max}$ and $T_{\min}$ such that $T_{\min} \leq T_n \leq T_{\max}$. For perfect reconstruction, we should have $T_{\max} = \beta T_{\nyq}$, where $0<\beta\leq 1 $.

To determine $T_{\max}$,  we would like the signal to be as close as possible to the bias so that the input to the integrator is comparatively low, which results in a large firing time. Hence, we proceed by considering the green and orange areas as depicted in Fig~\ref{fig:time_analysis}. The area of the parallelogram enclosed by these can be written as
\begin{align}
\Delta &= \mu  \text{$T_{\max}$}.
\label{eq:LB_tmax}
\end{align}
Then, for perfect reconstruction, we have that
\begin{align}
\mu = \frac{\Delta}{\beta T_{\nyq}}.  
\label{eq:cc}
\end{align}

Similarly, to calculate $T_{\min}$, we consider the scenario when the input to the integrator rapidly increases, as this results in the threshold being crossed quickly. Given our choice of bias, this happens when the signal increases at the maximum rate, which is along the slope $\epsilon$. Hence, we consider the areas depicted in purple and green in Fig.~\ref{fig:time_analysis}, and equate them to the threshold $\Delta$ as 
\begin{align}
\Delta &= \frac{1}{2} \epsilon T_{\min} \left( \epsilon T_{n-1} + \epsilon T_{n-1} + 2 T_{\min} \epsilon \right) + \mu  T_{\min}.
\label{eq:LB_tmin}
\end{align}
Solving the quadratic equation in \eqref{eq:LB_tmin} we have
\begin{align}
T_{\min}
\;=\;
\frac{ -\big(\epsilon T_{n-1} + \mu\big) + \sqrt{\big(\epsilon T_{n-1} + \mu\big)^2 + 4\epsilon\Delta } }{2\epsilon}.
\label{eq:tmin_quad}
\end{align}
Note that the right-hand side of \eqref{eq:tmin_quad} is a monotone decreasing function of the previous interval \(T_{n-1}\); hence the smallest possible \(T_{\min}\) occurs when the previous interval is maximal: \(T_{n-1} = \beta T_{\nyq}\). 
We now enforce the condition \(t_{\min} = \alpha T_{\nyq}\). Substituting these into \eqref{eq:tmin_quad} and rearranging yields the relation
\begin{align}
\Delta
&= \alpha\epsilon T_{\nyq}^2(\beta + \alpha) + \alpha T_{\nyq}\,\mu.
\label{eq:delta_intermediate}
\end{align}
Using \eqref{eq:cc} to eliminate \(\mu\) we obtain
\[
\Delta
= \alpha\epsilon T_{\nyq}^2(\beta + \alpha) + \frac{\alpha}{\beta}\,\Delta.
\]
Solving for \(\Delta\) yields the closed-form
\begin{align}
 \displaystyle
\Delta \;=\; \frac{\alpha\beta\,\epsilon\,T_{\nyq}^2(\beta + \alpha)}{\beta - \alpha}
\label{eq:delta_final}
\end{align}
and therefore, using \eqref{eq:cc},
\begin{align}
 \displaystyle
\mu \;=\; \frac{\Delta}{\beta T_{\nyq}}
\;=\; \frac{\alpha\,\epsilon\,T_{\nyq}(\beta + \alpha)}{\beta - \alpha}.
\label{eq:c_final}
\end{align}

With \(\Delta\) and \(\mu\) chosen according to \eqref{eq:delta_final}--\eqref{eq:c_final}, the worst-case minimal interval equals \(T_{\min}=\alpha T_{\nyq}\) and the maximal interval equals \(T_{\max}=\beta T_{\nyq}\); which completes the proof.
\end{proof}

The result shows that $\alpha$ and $\beta$ can be freely selected (subject to $\alpha<\beta\le1$), enabling explicit control of the oversampling ratio. Unlike the variable-bias IF-TEM (VB-IF-TEM), the LB-IF-TEM imposes no upper limit on achievable oversampling.

Moreover, event-driven sampling excels for bursty signals but over-samples bandlimited signals in high-amplitude regions. By tracking the signal with a linear bias, the integrator sees an almost constant input, yielding near-uniform, amplitude-insensitive firings. This design offers explicit control of sensitivity (via $\Delta,\mu$) and removes VB-IF-TEM’s oversampling limits—better matching the uniformly distributed information in bandlimited signals and enabling more efficient low-bit NUQ.

Our motivation in designing the LB-IF-TEM was to address the limited oversampling capability of the VB-IF-TEM, as demonstrated in Theorem~\ref{thm:linear-bias-intervals}, and to achieve more concentrated firing times that enable efficient deployment of an NUQ scheme. To this end, we compare the proposed LB-IF-TEM with the conventional IF-TEM (with constant bias)~\cite{lazar2003time} and the VB-IF-TEM~\cite{VB-IFTEM}. Table~\ref{tab:tem_comparison_combined} summarizes the parameter expressions for all three time-encoding mechanisms, along with the corresponding minimum and maximum firing intervals and the interval range $T_{\text{range}} = T_{\max} - T_{\min}$. Here, $\Delta_{\mathrm{c}}$ and $b$ denote the threshold and bias for the conventional IF-TEM, respectively, while $\Delta_{\mathrm{v}}$ and $\Delta_{l}$ represent the thresholds in the VB-IF-TEM and LB-IF-TEM. For a more direct comparison, the table also lists specific parameter values used in our simulations. The parameters are selected such that all three schemes achieve a normalized mean-squared error (NMSE) of $-50,\mathrm{dB}$ in reconstructing bandlimited signals, while maintaining $T_{\max} = T_{\nyq}$.

\begin{figure}[!t]
    \centering
    \includegraphics[width=3.3 in]{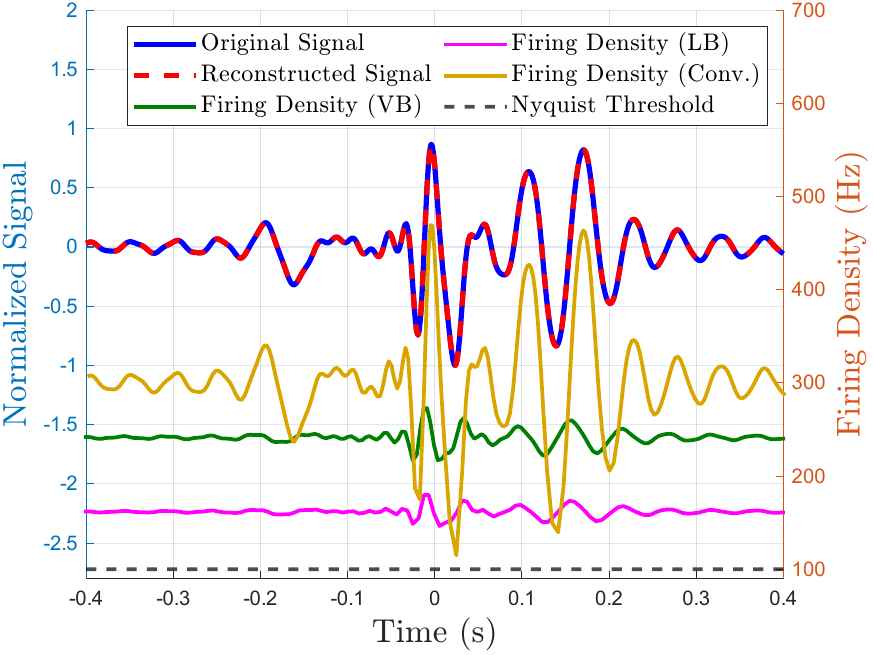}
    \caption{Firing rates for the different TEM variants}
    \label{fig:difffiring}
\end{figure}

\begin{table*}[!t]
\caption{Comparison of $T_{\min}$, $T_{\max}$, and $T_{\text{range}}$ across TEMs (General and Numerical Values)}
\centering
\setlength{\tabcolsep}{6.5pt}
\renewcommand{\arraystretch}{1.25}
\begin{tabular}{lccc|ccc}
\toprule
\multirow{2}{*}{{TEM}} 
& \multicolumn{3}{c|}{{General Expressions}} 
& \multicolumn{3}{c}{{Numerical Values (millisec.)}} \\
\cmidrule(lr){2-4} \cmidrule(l){5-7}
& ${T_{\max}}$ 
& ${T_{\min}}$ 
& ${T_{\text{range}}}$
& ${T_{\max}}$ 
& ${T_{\min}}$ 
& ${T_{\text{range}}}$ \\
\midrule
Conv 
& $\frac{\Delta_{\text{c}}}{b-c}$ 
& $\frac{\Delta_{\text{c}}}{b+c}$ 
& $\frac{2c\Delta_{\text{c}}}{b^2-c^2}$ 
& 10 & 1.2 & 8.8 \\
VB   
& $\sqrt{\frac{2\Delta_{\text{v}}}{\epsilon}}$        
& $(\sqrt{5}-2)\sqrt{\frac{2\Delta_{\text{v}}}{\epsilon}}$ 
& $(3-\sqrt{5})\sqrt{\frac{2\Delta_{\text{v}}}{\epsilon}}$ 
& 10 & 2.4 & 7.6 \\
LB           
& $\frac{\Delta_{l}}{\mu}$  
& $\frac{-(\epsilon\Delta_{l}+\mu^2)+\sqrt{(\epsilon\Delta_{l} +\mu^2)^2 +4\epsilon\Delta_{l}\mu^2}}{2\epsilon\mu}$   
& $\frac{\Delta_{l}}{\mu} - \frac{-(\epsilon\Delta_{l}+\mu^2)+\sqrt{(\epsilon\Delta_{l} +\mu^2)^2 +4\epsilon\Delta_{l}\mu^2}}{2\epsilon\mu}$ 
& 10 & 4.1 & 5.9 \\
\bottomrule
\end{tabular}
\\[4pt]
\footnotesize
Parameters: $\Delta_{\text{c}}=0.02$, $b=1.5$, $\Delta_{\text{v}}=0.0157$, $\Delta_{{l}}=0.0314$, $\mu=3.14$
\label{tab:tem_comparison_combined}
\end{table*}
From Table~\ref{tab:tem_comparison_combined}, it is evident that the LB-IF-TEM produces a narrower range of firing intervals compared to the other methods. To illustrate the firing behavior with signal variation, Fig.~\ref{fig:difffiring} shows a bandlimited signal (blue), its reconstruction using the LB-IF-TEM (red), and the corresponding firing densities (computed as $1/T_n$). The densities in the LB-IF-TEM are concentrated within approximately $10~\text{Hz}$, whereas the spreads are about $20~\text{Hz}$ and $50~\text{Hz}$ for the VB-IF-TEM and conventional schemes, respectively.

While limiting the variation in firing intervals can be disadvantageous for spiky or highly localized signals that benefit from adaptive timing, it is advantageous for smoothly varying bandlimited signals, where information is more uniformly distributed in time. This validates our design intuition of concentrating firings to improve quantization efficiency. In the next section, we analyze the simulation results to quantify the resulting gain in communication cost.

\section{Experimental Results}
\label{sec:ER}

We evaluate the proposed LB-IF-TEM using bandlimited signals with bandwidth $\Omega_0 = 100\pi$ and maximum amplitude $c=1$, defined over the interval $[-0.45, 0.45]$ seconds. Under Shannon–Nyquist sampling, perfect reconstruction would require $2\times (\Omega_0 / 2\pi)\times \text{(time interval)} = 90$ samples. To study quantization performance, the parameters of the conventional, VB-, and LB-IF-TEMs are tuned to achieve an NMSE of $-50$~dB. The corresponding numbers of firings required are $266$, $214$, and $144$, respectively.

To design the NUQ, we first construct histograms of the firing intervals (over 100 randomly generated signals) and generate codebooks using both the Max–Lloyd algorithm \cite{max_lloyd} and a power-law-based companding strategy \cite{companding}. For each TEM framework, the NUQ configuration yielding the lower average NMSE (over 100 signals) is retained.

\begin{figure}[!t]
    \centering
    \includegraphics[width= 3.2in]{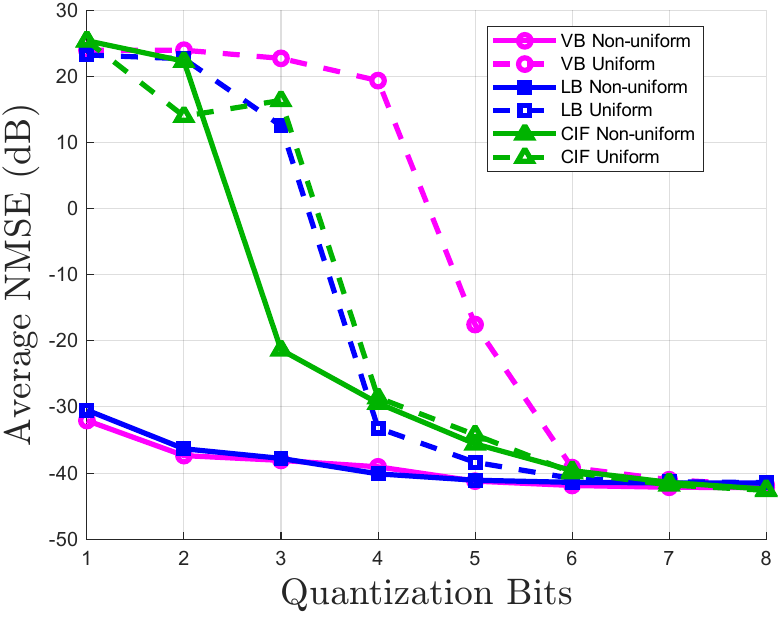}
    \caption{Comparison of NMSE versus bit budget for different IF-TEM schemes under UQ/NUQ.}
    \label{fig:main_error_plot}
\end{figure}

Fig.~\ref{fig:main_error_plot} compares the NMSEs for UQ and NUQ as a function of bit rate. The NUQ consistently achieves lower error than the UQ in both VB-IF-TEM and LB-IF-TEM, particularly in the low-bit regime. In contrast, the performance gain is less pronounced for the conventional IF-TEM (labelled as CIF), as its constant bias leads to greater variance in firing intervals. Both VB-IF-TEM and LB-IF-TEM achieve comparable reconstruction accuracy under NUQ; however, the LB-IF-TEM requires roughly $1.5\times$ fewer firings to reach the same NMSE. 

These results confirm that the proposed LB-IF-TEM achieves lower distortion and reduced communication cost by effectively localizing the firing intervals and exploiting non-uniform quantization more efficiently.

\section{Conclusion}
\label{sec:C}

This letter presented the linear-bias IF-TEM, along with theoretical bounds on its achievable oversampling range. Unlike the variable-bias IF-TEM, the proposed method imposes no intrinsic constraints from bias dynamics. By designing a bias that closely tracks the signal, the LB-IF-TEM maintains a nearly constant integrator input, resulting in low variance of firing intervals. 

Exploiting this property, we employed a NUQ optimized using the empirical histogram of firing intervals. Experimental results confirm that the LB-IF-TEM achieves superior rate–distortion performance and reduced communication cost compared to conventional and VB-based IF-TEMs. Notably, even with 1-bit quantization and an oversampling rate of only 1.6$\times$ the Nyquist rate, the proposed method attains significantly lower reconstruction error.

Overall, the LB-IF-TEM offers a low-power, communication-efficient, and scalable framework for time-based signal encoding, applicable not only to bursty signals but also to smoothly varying bandlimited signals with uniformly distributed information.

\bibliographystyle{IEEEtran}
\bibliography{refs3}

\end{document}